\documentclass{aa}
\usepackage{natbib}
\usepackage{graphicx}
\usepackage{latexsym}
\usepackage{txfonts}
\newcommand{\nc}{\newcommand}  
\nc{\teff}{$T_{\rm eff}$\,}  
\nc{\logg}{log\,$g$\,}  
\nc{\kms}{\,${\rm km\,s}^{-1}$\,}  
\nc{\mic}{$\xi_{\rm t}$\,}
\nc{\feh}{\,${\rm [Fe/H]}$\,}
\nc{\csthree}{CS~31080$-$095}
\nc{\cstwo}{CS~22958$-$042}
\nc{\cstwonine}{CS~29528$-$041}
\nc{\cratio}{$\rm^{12}$C/$^{13}$C}
\newcommand\nodata{ ~$\cdots$~ }
\newcommand {\msun}{\mbox{M$_\odot$}}
\newcommand{\czw}{$^{12}$C} 

\begin{document}

\title{First stars X. The nature of three unevolved Carbon-Enhanced Metal-Poor 
stars
\thanks {Based on observations made with the ESO Very Large Telescope at 
Paranal Observatory, Chile (Large programme ``First Stars'', 
ID 165.N-0276(A); P.I.: R. Cayrel)}
}
\author{
T. Sivarani\inst{1} \and
T.~C. Beers\inst {1} \and
P. Bonifacio\inst {2,3,4} \and
P. Molaro\inst {4,3} \and
R. Cayrel\inst {3} \and
F. Herwig\inst {5} \and 
M. Spite\inst {3} \and
F. Spite \inst {3} \and
B. Plez\inst {6} \and
J. Andersen\inst {7,8} \and
B. Barbuy\inst {9} \and
E. Depagne\inst{10} \and
V. Hill\inst {3} \and 
P. Fran\c cois\inst{3} \and
B. Nordstr\"om\inst {11,7} \and 
F. Primas\inst{12} 
}

\offprints{T. Sivarani ({\it thirupathi@pa.msu.edu})}

\institute{
 Department of Physics \& Astronomy, CSCE: Center for the Study of Cosmic 
Evolution,
 and JINA: Joint Institute for Nuclear Astrophysics, Michigan State University, 
East Lansing, MI 48824, USA
\and 
CIFIST Marie Curie Excellence Team
\and  
  Observatoire de Paris, GEPI, F-92195 Meudon Cedex, France
\and 
     Istituto Nazionale di Astrofisica - Osservatorio 
     Astronomico di Trieste, Via Tiepolo 11, I-34131 Trieste, Italy 
\and
Los Alamos National Laboratory (LANL), Los Alamos, New Mexico, NM 87545
\and   
  GRAAL, Universit\'e de Montpellier II, F-34095 Montpellier Cedex 05,
           France
\and  
        The Niels Bohr Institute, Astronomy, Juliane Maries Vej 30,
                   DK-2100 Copenhagen, Denmark
\and  
    Nordic Optical Telescope Scientific Association, Apartado 474,
    ES-38 700 Santa Cruz de La Palma, Spain
\and  
    Universidade de Sao Paulo, Departamento de Astronomia,
Rua do Matao 1226, 05508-900 Sao Paulo, Brazil
\and
  European Southern Observatory (ESO), 3107 Alonso de Cordova, Vitacura, 
  Casilla 19001, Santiago 19, Chile
\and 
Lund Observatory, Box 43, S-221 00 Lund, Sweden
\and 
European Southern Observatory (ESO), Karl-Schwarschild-Str. 2,
              D-85749 Garching b. M\"unchen, Germany
}

\authorrunning{T. Sivarani et al.}
\titlerunning{First stars X.}
\date{Received xxx; Accepted 8 July 2006}

\abstract
{On the order of 20\% of the very metal-poor stars in the Galaxy exhibit large carbon 
enhancements. It is important to establish which astrophysical sites and 
processes are responsible for the elemental abundance patterns of this early 
Galactic population.
}
{We seek to understand the nature of the progenitors of three main-sequence
turnoff Carbon-Enhanced Metal-Poor (CEMP) stars, \csthree, \cstwo, and 
\cstwonine, based on a detailed abundance analysis.
}
{From high-resolution VLT/UVES spectra ($R\sim 43,000$), we determine 
abundances or upper limits for Li, C, N, O, and other important elements, as
well as $^{12}$C/$^{13}$C isotopic ratios.
} 
{All three stars have $\rm -3.30 \le [Fe/H] \le -2.85$ and moderate to high 
CNO abundances. 
\cstwo\ is one of the most carbon-rich CEMP stars known ([C/Fe] = +3.2), while
\cstwonine\ (one of the few N-enhanced metal-poor stars known) 
is one of the most nitrogen rich ([N/Fe] = +3.0). 
Oxygen is very high in \csthree\ ([O/Fe] = +2.35) and in \cstwo\ ([O/Fe] = +1.35).
All three stars exhibit [Sr/Fe] $< 0$; Ba is not detected in \cstwo\ 
([Ba/Fe] $< -0.53$), but it is moderately enhanced ([Ba/Fe] $\sim 1$) in the 
other two stars. \cstwo\ displays one of the largest sodium overabundances
yet found in CEMP stars ([Na/Fe] = +2.8).
\cstwo\ has $^{12}$C/$^{13}$C = 9, similar to most other CEMP stars without
enhanced neutron-capture elements, while $^{12}$C/$^{13}$C $\ge 40$ 
in \csthree. 
\csthree\ and \cstwonine\ have A(Li) $\sim 1.7$, below the Spite Plateau, 
while Li is not detected in \cstwo.
}
{\cstwo\ is a CEMP-no star, but the other two stars are in no known class of 
CEMP star and thus either constitute a new class or are a link between 
the CEMP-no and CEMP-s classes, adding complexity to the abundance patterns 
for CEMP stars. We interpret the abundance patterns in our stars to imply
that current models for the presumed AGB binary progenitors lack an 
extra-mixing process, similar to those apparently operating in RGB stars.
}
\maketitle

\section{Introduction}

\begin{figure}[th]
\rotatebox{0}{\resizebox{8.5cm}{!}{\includegraphics{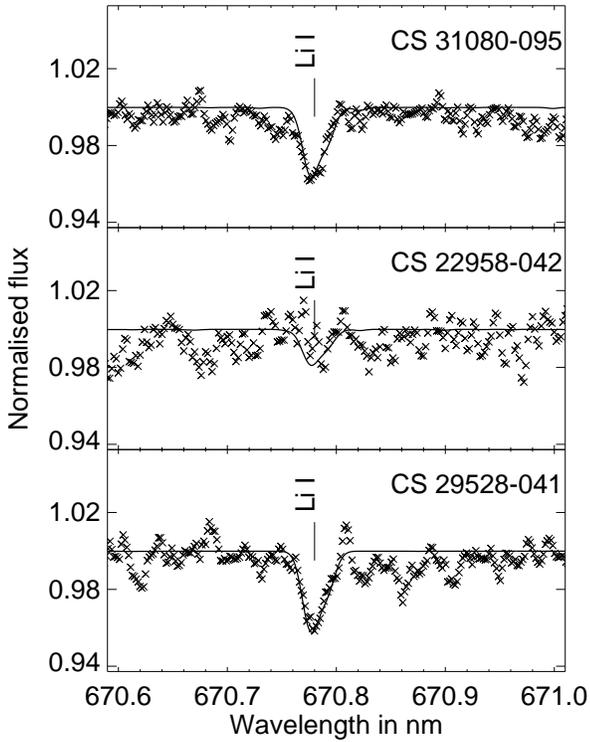}}}
\caption{Observed (crosses) and synthetic spectra (lines) of \csthree, \cstwo,
and \cstwonine\ in the region of the Li~I 670.7 nm doublet. Li is not detected 
in \cstwo.}
\end{figure}

A considerable fraction (20-25\%) of the very metal-poor stars in the Galaxy 
([Fe/H] $\le -2.0$) are strongly enhanced in carbon and often in nitrogen as 
well ([C,N/Fe] $>$ 1.0; Beers \& Christlieb 2005; Lucatello et al. 2006). It
appears that multiple nucleosynthetic pathways for the production of C in
low-metallicity stars must be invoked, almost certainly involving different
astrophysical sites. Most stars in the range $-2.6 \le 
\rm{[Fe/H]} \le -2.0$ appear to be associated with C production in 
Asymptotic Giant Branch (AGB) companions that have transferred material 
to the star that is now observed (Aoki et al. 2006a). Even 
lower-metallicity stars, especially those with [Fe/H] $< -3.0$, may contain 
C produced by massive, 
rapidly-rotating stars with [Fe/H] $< -6.0$; see, e.g.,
Hirschi et al. (2006); Karlsson et al. (2006); Piau et al. (2006). In order 
to better understand the role of these various possible mechanisms, detailed 
investigations are needed of the elemental abundances in 
Carbon-Enhanced Metal-Poor (CEMP) stars ([C/Fe] $> +1.0$; see Beers \& 
Christlieb 2005).  

Recent high-resolution spectroscopic studies of CEMP stars selected from the two
large objective-prism surveys for low-metallicity stars, the HK 
\citep{BPS1985,BPS1992,beers99} and Hamburg-ESO surveys \citep{HES}, 
have revealed a wide
variety of elemental abundance patterns (e.g, \citealt{norris1997b,norris1997c,
bonifacio1998,aoki2002a, aoki2002b, aoki2002c}; Barklem et al. 2005; Beers \&
Christlieb 2005). Many of these stars (roughly 70\% - 80\%; Aoki et
al. 2003, 2006a) exhibit strong s-process enhancements (CEMP-s), and are thought to be
the result of AGB mass transfer to the presently observed companion (e.g.,
\citealt{aoki2002c, lucatello2003, sivarani2004, barbuy2005}). 

The observed radial-velocity variations among the CEMP-s stars indicate that 
most, if not all, could be members of binary systems (\citealt{tsang2004,
lucatello2005}). Even within this group of CEMP-s stars, the observed s-process 
patterns are not identical, however. For example, the scatter in their observed 
Sr, Ba, Pb, and C abundances at a given [Fe/H] is large (e.g., 
\citealt{aoki2002c, sivarani2004, barbuy2005}). 
This diversity presumably arises because of the range in masses of the star
that underwent AGB evolution (see Herwig 2005 for a recent review), as well as
due to details of internal mixing and possible variations in the level of dilution 
of the material passed to the star we now observe.

\begin{figure}[th]
\rotatebox{0}{\resizebox{8.5cm}{!}{\includegraphics{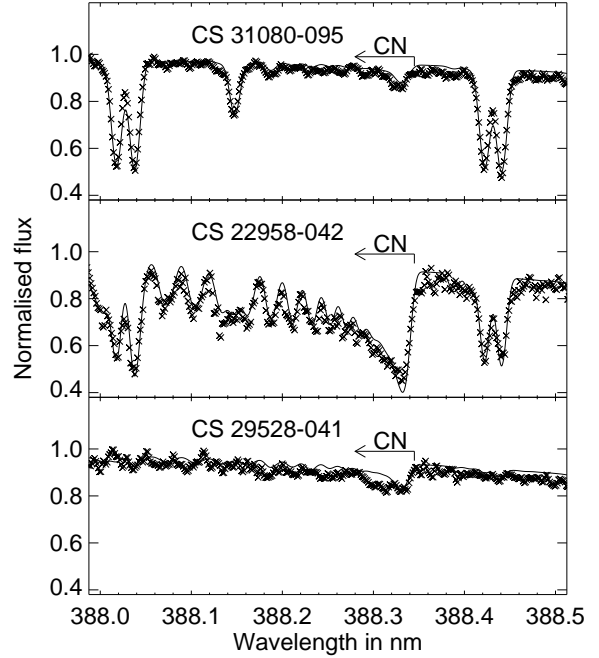}}}
\caption{Observed and synthetic spectra of our stars around the 388.3 nm CN
band (symbols as in Fig. 1). 
The strong CN bands in \cstwo\ indicate a larger N excess than in \csthree. 
In \cstwonine\ the CN bands are weak, but the derived [N/H] is high because 
[C/H] is low and \teff\ is high. The 336 nm NH bands are also detected in 
\cstwonine\ and \cstwo\ (not shown). } 
\end{figure}

\begin{table*}[ht]
\caption{Log of observations}
\begin{tabular}{lclllll}

\hline
  Star         & Coordinates     &       Date     &  MJD-24000.5   & Exposure  & 
Radial Velocity  & S/N\\
&(2000)&y/m/d&&(sec)&\kms&at 430 nm\\
\hline
                   
CS~31080-095   &  04:44:21.8 $-$45:13:57 &   2000-10-19   &  51836.2572450 & 
2400& $-$43.64 &93 \\
               &                         &   2000-10-19   &  51836.2864673 & 
2400& $-$43.30 &72 \\
               &                         &   2001-10-19   &  52201.3331806 & 
3600& $-$43.76 &82 \\
               &                         &   2001-10-20   &  52202.3388291 & 
3000& $-$43.76 &52 \\
               &                         &   2001-11-06   &  52219.2454194 & 
3600& $-$43.66 &92 \\
               &                         &   2001-11-09   &  52222.2693614 & 
1000& $-$43.34 &48 \\
&&&&&\\
CS~22958-042   &  02:01:07.5 $-$57:17:07 &   2001-11-05   &  52218.1900217 & 
3600& $+$165.21& 48 \\
               &                         &   2001-11-05   &  52218.2360571 & 
3600& $+$162.51& 35 \\
&&&&&\\
CS~29528-041   &  02:29:25.1 $-$18:13:30 &   2001-09-06   &  52158.3486878 & 
5000& $-$285.1 & 50\\
               &                         &   2001-10-21   &  52203.2220389 & 
3260& $-$285.4 & 33 \\  
               &                         &   2001-11-07   &  52220.2338976 & 
3600& $-$285.0 & 37 \\
\hline
\end{tabular}
\end{table*}

\begin{figure}[th]
\rotatebox{0}{\resizebox{8.5cm}{!}{\includegraphics{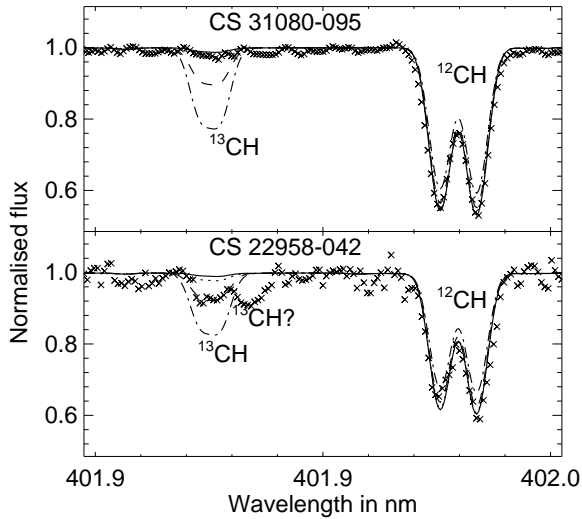}}}
\caption{Observed spectra (crosses) for the two stars with meaningful 
measurements or limits on the $^{12}$C/$^{13}$C ratio. Solid, dotted, 
dashed, and dash-dotted lines show our synthetic spectra for
$^{12}$C/$^{13}$C = 99, 40, 9, and 3, respectively. For \cstwonine\, all 
$^{12}$C/$^{13}$C ratios between 3 and 99 are within a 2 $\sigma$ 
detection limit; hence, no $^{12}$C/$^{13}$C is given for this star.}
\end{figure}

\begin{table}[h]
\caption{Photometry and adopted atmospheric parameters}
\begin{tabular}{lrrr}
\hline
Parameter          & CS~31080-095   & CS~22958-042  & CS~29528-041\\
\hline
$V$            &  12.989       & 14.516       & 14.60\\
$B-V$          &  0.521        & 0.479        & 0.42\\
$U-B$          & $-$0.291      &  \nodata     & $-$0.19\\
$V-R$          &  0.317        & 0.294        & \nodata \\
$V-I$          &  0.617        & 0.614        & \nodata \\
$V-K$          &  1.397        & 1.303        & 1.292\\
$J-H$          &  0.272        & 0.227        & 0.314 \\
$J-K$          &  0.350        & 0.302        & 0.339 \\
               &               &              & \\
E(B$-$V)       &  0.009        & 0.025        & 0.030 \\
               &               &              & \\
\teff($B-V$)$_{0}$   &  5671         & 5897         & 6378 \\
\teff($V-R$)$_{0}$   &  5964         & 6221         &\nodata \\
\teff($V-I$)$_{0}$   &  6020         & 6146         &\nodata \\
\teff($R-I$)$_{0}$   &  6224         & 6135         &\nodata \\
\teff($V-K$)$_{0}$   &  6077         & 6344         & 6199 \\
\teff($J-H$)$_{0}$   &  5940         & 6277         & 5947 \\
\teff($J-K$)$_{0}$   &  5822         & 6156         & 6266 \\
               &               &              & \\
Adopted: & & & \\
\teff(Fe lines)      & 6050          & 6250         & 6150 \\
\logg                & 4.50          & 3.50         & 4.00 \\
\feh                 & $-$2.85       &$-$2.85       & $-$3.30 \\
\mic(\kms)           & 1.0    & 1.5          & 1.3 \\
\hline                                     
\end{tabular}                              
\end{table}                                

Other CEMP stars exhibit r-process enhancements (e.g.,
CS~22892-052; Sneden et al. 2003); Lucatello (private communication) reports 
several additional likely CEMP-r stars from an analysis of the
HERES sample of Barklem et al. (2005). Another class of carbon-enhanced
metal-poor stars are the CEMP-r/s stars, which exhibit the presence of
neutron-capture elements associated with {\it both} the r- and s-process (e.g.,
CS~22948-027, CS~29497-034, Hill et al. 2000 and Barbuy et al. 2005; CS
29526-110, CS~22898-027, CS~31062-012, Aoki et al. 2002c; HE~2148-1247, Cohen et
al. 2003; CS~29497-030, Ivans et al. 2005). 

Most CEMP-r/s stars also
exhibit very high Pb abundances (the so-called ``lead stars," see Aoki et al.
2002c and van Eck et al. 2003). The discovery of CS~22957-027
\citep{norris1997c,bonifacio1998} and a number of similar stars which 
exhibit {\it no} enhancements of neutron-capture elements
\citep{aoki2002a,aoki2002b,depagne,2005astro.ph..1535P,2005Natur.434..871F}, 
led Beers \& Christlieb (2005) to introduce the CEMP-no class of stars. 

As shown below, at least one of the three stars studied here is
classified as a CEMP-no star. The other two stars in our study exhibit
intermediate Ba excesses ([Ba/Fe] $>$ 0), but no other strong
s-process-element enrichment. Thus, they fail the Beers \& Christlieb (2005)
criterion for the Ba ratios in CEMP-no stars (which requires [Ba/Fe] $<$ 0.0),
but they may well be related to this class. 

The origin of CEMP-no stars remains
unclear at present. They could, for example, be the result of pollution by AGB
stars which, for some reason, have been unable to activate the s-process, but
have still dredged up significant amounts of carbon to their outer atmospheres.
Alternatively, the high CNO abundances in these stars could be the products of
SNe; such models have been proposed at least in the case of HE~0107-5240
\citep{2002Natur.419..904C,2003Natur.422..834B} and HE~1327-2326
\citep{2005Natur.434..871F,2005Sci...309..451I}.

Measurements of C, N, and O abundances provide crucial
information on the nucleosynthetic and mixing history of the progenitors of CEMP
stars. The $^{12}$C/$^{13}$C ratio is a sensitive indicator of the mixing
processes experienced by carbon-enhanced stars; Li provides another
valuable diagnostic tool. Lithium is not detected in most CEMP stars, but is
sometimes found at a level below the Spite Plateau
(\citealt{norris1997a, aoki2002b}); both observations suggest that some
mechanism induced strong mixing and Li burning in the past. However, 
at least two known CEMP stars, LP~706-7 \citep{norris1997b} and CS~22898-027 
\citep{TB1992} exhibit Li abundances corresponding to the Spite Plateau. 
Perhaps surprisingly, both of these stars are neutron-capture-element-rich 
stars (\citealt{norris1997b, aoki2002c}), although significant radial-velocity 
variation has yet to be found in either star.

In this paper we present a detailed abundance analysis for three
main-sequence turnoff CEMP stars, \csthree, \cstwo, and \cstwonine, based on
high-resolution, high S/N observations obtained with the ESO VLT/UVES.  
Main-sequence turnoff CEMP stars are of particular interest because their 
present evolutionary stage precludes the significant internal mixing that
is expected for cooler CEMP stars, hence their atmospheres better preserve
the nucleosynthetic signatures of their progenitors.   

We describe our observations in Sect. 2. Details of our abundance analysis
procedures are reported in Sect. 3; in Sect. 4 we present the results, 
and in Sect. 5 we examine possible scenarios for the nucleosynthesis 
histories of these objects and compare them with the properties of 
previously observed CEMP stars. Our conclusions are presented in Sect. 6.

Throughout this paper, the abbreviations VMP, EMP, UMP, HMP, and MMP (Very, 
Extremely, Ultra, Hyper, and Mega Metal Poor) denote stars with [Fe/H] 
below $-$2.0, $-$3.0, $-$4.0, $-$5.0, and $–$6.0, respectively (Beers \& Christlieb 2005).

\section{Observations and data reduction}

Our spectroscopic data were obtained at the VLT-Kuyen 8.2m telescope and 
UVES spectrograph (Dekker et al. 2000) at a resolving power of $R = 43000$.
The coordinates of our program stars and the log of observations are listed in 
Table 1. Non-standard grating settings were used, both with dichroic \#1: 
396+573, 396+850 (central wavelengths in nm in the blue and red arm,
respectively). The central wavelengths in the red arm were chosen so 
that both settings would cover the Li doublet, effectively doubling the 
integration time on this important line. 

The data were reduced using the UVES context within MIDAS, which includes bias
subtraction, flat fielding, wavelength calibration, and the merging of echelle
orders. The continuum normalisation was performed with IRAF (using a cubic
spline function) for the merged spectra. For a few lines that were either very
weak, or of particular interest, we have used the spectra of the individual
orders without merging them. Balmer-line profiles were checked using both
single-order and merged spectra. Equivalent width measurements for unblended
lines were obtained by fitting gaussian profiles, using the genetic algorithm
code described by \citet{francois}. 

Radial velocities (listed in Table 1) were measured from unblended lines in 
the range 380-450 nm. Multiple-epoch observations of \csthree\ (over a range 
of 386 days) and of \cstwonine\ (over a range of 45 days) 
do not reveal any detectable radial-velocity variations, while quite large (3 km
s$^{-1}$) variations are found in \cstwo\ over only one hour. If
this velocity variation is correct, \cstwo\ may be another example of a very
short-period binary similar to HE~0024-2523, as reported previously by Lucatello
et al. (2003).

$UBV$ photometry for \cstwonine\ is available from Norris, Ryan, \& Beers 
(1999).  For the other two stars, we obtained broadband photometry with the 
Danish 1.5m telescope at ESO, Chile (Beers et al., in preparation). Near
infrared JHK photometry for all three stars was available from the 2MASS catalog
(Skrutskie et al. 2006). Estimates of interstellar reddening along the line of sight
to each star were obtained from Schlegel, Finkbeiner, \& Davis (1998). We also 
adopt Table 6 from Schlegel, Finkbeiner, \& Davis (1998) for the relative 
extinctions of various bandpasses. The photometric data and derived 
parameters are listed in Table 2.  

All our program stars have intermediate-band Str\"omgren photometry
by Schuster et al. (2005), who estimated metallicities and photometric 
classifications as follows: \csthree: MS (CH), \cstwo: TO
(CH), and \cstwonine: TO, respectively; these classifications are
consistent with our derived surface gravities, as described below.

\section{Analysis}

For a first estimate of the \teff\ for our program stars, we employed the 
various colours with the \citet{alonso} calibrations of \teff\ for dwarf 
stars. The results are listed in Table 2. Transformations between the 
different photometric systems necessary for use of the Alonso et al. 
calibrations were carried out as described in \citet{sivarani2004}.
With the exception of temperatures based on the $B-V$ colour, which is 
susceptible to the effect of molecular carbon absorption on the $B$-band 
flux, the colour-based estimates of \teff agree to within 125-200 K.

As a likely better approach to estimating \teff we have 
also used the Fe~I excitation equilibrium.  The
effective temperatures derived from this procedure, also listed in Table 2, fall
within 100K of the average of the colour-based estimates of \teff (not including
the estimate based on $B-V$). In particular, they agree very well (within
50K) with the temperature from $V-K$, which should be superior due
to the large baseline and the fact that both the $V$ and $K$ bands
are relatively free of molecular carbon features. We adopt the spectroscopic 
\teff estimates for the remainder of our analysis. 

The surface gravity, log $g$, was derived from the 
Fe~I/Fe~II ionisation equilibrium, and the microturbulence was determined 
by requiring that the abundances from the Fe~I lines be
independent of equivalent width.  The gravity we obtain for CS~31080-095 
(log $g$ = 4.50) is consistent with the dwarf classification
by Schuster et al. (2005), and our results for CS~22958-042 (log $g$ = 3.50) 
and CS~29528-041 (log $g$ = 4.00) also correspond to 
main-sequence TO stars. The slightly lower gravity of CS~22958-042 may indicate 
that it is a subgiant that has evolved slightly away from the turnoff.

\begin{table*}[th]
\caption{Elemental abundance and isotopic ratio measurements for three 
 CEMP stars}
\begin{tabular}{lrrrrrrr}
\hline
Element &  \multicolumn{2}{c}{\csthree} & 
\multicolumn{2}{c}{\cstwo}&\multicolumn{2}{c} {CS~29528-041}&  Sun \\
        &  \multicolumn{1}{c}{log($\epsilon$)} & \multicolumn{1}{c}{[X/Fe]} 
        &  \multicolumn{1}{c}{log($\epsilon$)} & \multicolumn{1}{c}{[X/Fe]} 
        &  \multicolumn{1}{c}{log($\epsilon$)} & \multicolumn{1}{c}{[X/Fe]} 
        &  log($\epsilon$) \\
\hline
Li      & 1.73$\pm$0.05     & \nodata         &  $<$0.6            & \nodata           
& 1.71$\pm$0.05  & \nodata          & 1.10 \\
C(CH)   & 8.25$\pm$0.01     & +2.69 $\pm$0.14 &  8.70$\pm$0.01     & 
+3.15$\pm$0.12    & 6.70$\pm$0.08  & +1.59$\pm$0.15   & 8.41 \\
C(C2)   & 8.35$\pm$0.01     & +2.79 $\pm$0.14 &  8.70$\pm$0.01     & 
+3.15$\pm$0.12    &\nodata         & \nodata          & 8.41 \\
N(NH)   &                   & \nodata         &  7.05$\pm$0.02     & 
+2.10$\pm$0.12    & 7.50$\pm$0.1   & +3.00$\pm$0.15   & 7.80 \\
N(CN)   & 5.65$\pm$0.01     & +0.70 $\pm$0.11 &  7.10$\pm$0.01     & 
+2.15$\pm$0.11    & 7.57$\pm$0.06  & +3.07$\pm$0.13   & 7.80 \\
O(LTE)  & 8.17$\pm$0.05     & +2.35 $\pm$0.12 &  7.17$\pm$0.05     & 
+1.35$\pm$0.11    & $<$6.77        &$<$+1.40$\pm$0.10 & 8.67 \\ 
O(NLTE) & 8.05$\pm$0.05     & +2.23 $\pm$0.12 &  6.94$\pm$0.05     & 
+1.12$\pm$0.11    & $<$6.77        &$<$+1.40$\pm$0.10 & 8.67 \\
Na(LTE) & 3.20$\pm$0.05     & $-$0.28$\pm$0.12&  6.30$\pm$0.10     & 
+2.82$\pm$0.15    & 4.23$\pm$0.05  & +1.20$\pm$0.11   & 6.33 \\
Na(NLTE)& 3.00$\pm$0.05     & $-$0.48$\pm$0.12&  6.10$\pm$0.10     & 
+2.62$\pm$0.15    & 4.02$\pm$0.05  & +1.00$\pm$0.11   & 6.33 \\
Mg I    & 5.38              & +0.65 $\pm$0.12 &  5.05$\pm$0.04     & 
+0.32$\pm$0.15    & 4.68$\pm$0.01  & +0.40$\pm$0.11   & 7.58 \\
Al(LTE) & 2.67              & $-$0.95$\pm$0.11&  2.77              & $-
$0.85$\pm$0.10  & 2.77           & $-$0.85$\pm$0.10 & 6.47 \\
Al(NLTE)& 3.32              & $-$0.30$\pm$0.11&  3.32              & $-
$0.20$\pm$0.10  & 3.32           & $-$0.20$\pm$0.10 & 6.47 \\
Si I    & 4.75              & +0.05 $\pm$0.15 &  4.85              & 
+0.15$\pm$0.10    & 4.05           & $-$0.20$\pm$0.10 & 7.55 \\
Ca I    & 3.68$\pm$0.05     & +0.17 $\pm$0.13 &  3.87$\pm$0.06     & 
+0.36$\pm$0.15    & 3.46$\pm$0.03  & +0.40$\pm$0.15   & 6.36 \\
Sc II   & 0.30$\pm$0.05     & $-$0.02$\pm$0.14&  0.37$\pm$0.05     & 
+0.05$\pm$0.11    & 0.13$\pm$0.03  & +0.26$\pm$0.15   & 3.17 \\
Ti I    & 2.59$\pm$0.03     & +0.42 $\pm$0.16 &  2.60              & 
+0.43$\pm$0.10    & 2.32$\pm$0.04  & +0.40$\pm$0.15   & 5.02 \\
Ti II   & 2.39$\pm$0.01     & +0.22 $\pm$0.11 &  2.36$\pm$0.07     & 
+0.19$\pm$0.14    & 2.12$\pm$0.007 & +0.40$\pm$0.10   & 5.02 \\
V  I    & 1.15$\pm$0.05     & +0.00 $\pm$0.12 &  1.10              & $-
$0.05$\pm$0.10  & 0.70           & +0.00$\pm$0.15   & 4.00 \\
Cr I    & 2.84$\pm$0.03     & +0.02 $\pm$0.12 &  2.67$\pm$0.04     & $-
$0.15$\pm$0.11  & 2.20$\pm$0.01  &$-$0.17$\pm$0.11  & 5.67 \\
Mn I    & 2.33$\pm$0.05     & $-$0.21$\pm$0.13&  2.33$\pm$0.05     & $-
$0.21$\pm$0.14  & 1.59$\pm$0.07  &$-$0.50$\pm$0.15  & 5.39 \\
Fe I    & 4.65$\pm$0.003    & $-$0.00$\pm$0.10&  4.65$\pm$0.04     & 
+0.00$\pm$0.10    & 4.20$\pm$0.002 &+0.00$\pm$0.10    & 7.50 \\
Fe II   & 4.66$\pm$0.02     & $-$0.05$\pm$0.11&  4.57$\pm$0.09     & $-
$0.08$\pm$0.12  & 4.20$\pm$0.02  &+0.00$\pm$0.10    & 7.50 \\
Co I    & 2.38$\pm$0.03     & +0.31 $\pm$0.11 &  2.27$\pm$0.14     & 
+0.20$\pm$0.18    & 1.62$\pm$0.04  &+0.00$\pm$0.15    & 4.92 \\
Ni I    & 3.49$\pm$0.03     & +0.09 $\pm$0.16 &  3.31$\pm$0.10     & $-
$0.09$\pm$0.16  & 2.95$\pm$0.01  &+0.00$\pm$0.12    & 6.25 \\
Zn I    & 2.33$\pm$0.08     & +0.58 $\pm$0.15 &  \nodata           &\nodata            
& \nodata        & \nodata  & 4.60\\
Sr II   & $-$0.29$\pm$0.05  & $-$0.41$\pm$0.12&  $-$0.53           & $-
$0.20$\pm$0.11  &$-$0.53         & $-$0.20$\pm$0.10 & 2.97 \\
Y  II   & $-$0.96$\pm$0.05  & $-$0.35$\pm$0.13&  \nodata           & \nodata           
&\nodata         &\nodata  & 2.24 \\
Ba II   & 0.05$\pm$0.05     & +0.77  $\pm$0.15&  $<$$-$1.40        & $<$$-
$0.53$\pm$0.16  &$-$0.23      & +0.97$\pm$0.10   & 2.13 \\
$^{12}$C/$^{13}$C & $>$ 40  &         &  9.0$\pm$2.0       & \nodata  &\nodata         
&\nodata         &\nodata    \\
\hline
\end{tabular}
\end{table*}

\section{Abundances}

We employ the same model atmospheres (\texttt{OSMARCS}; see Gustafsson 
et al. 2003 and references therein, with appropriate C and N enhancement), 
current version of the spectrum synthesis code (\texttt{turbospectrum};
\citealt{alvarez-plez}), and line lists as 
in previous papers of this series \citep{hill,depagne, francois}. We also 
use the CH molecular line list compiled by Plez (Plez \& Cohen 2005). The 
NH and C$_2$ molecular linelists are taken from the Kurucz database
(\texttt{http: //kurucz.harvard.edu/linelists/linesmol/}). 

Our derived elemental abundances are listed in Table 3. Errors in
log($\epsilon$) are assigned for stars with multiple lines, based on the errors
in the mean (i.e., $\sigma$(log($\epsilon))/\sqrt{(N)}$); no errors are listed
for elements in which only one line was measured. Total errors in [X/Fe] for 
each element are of the order of 0.10-0.15 dex, taking our adopted errors of 
100K in \teff\ and 0.5 dex in \logg\ into account. We adopt the solar abundances
of Asplund, Grevesse, \& Sauval (2005).

\cstwo\ has no detectable Ba ([Ba/Fe] $< -0.53$) (Table 3); it is 
thus classified as a CEMP-no star, according to
Beers \& Christlieb (2005). The other two stars exhibit moderate enhancements of
Ba ([Ba/Fe] = $+0.77$ and $+0.97$). These values of Ba enrichment fall below the
level suggested by Beers \& Christlieb for the CEMP-s classification ([Ba/Fe $>
+1.0)$, but they do satisfy the Ryan et al. (2005) criterion for Ba-rich stars, 
[Ba/Fe]$ > +0.50$, which is also adopted by Aoki et al. (2006a).
Note that the Sr abundances in these two stars are
quite low, [Sr/Fe] $= -0.41$ (\csthree) and [Sr/Fe] = $-0.20$ (\cstwonine).

\csthree\ and \cstwo\ are more C rich than N rich. In contrast, \cstwonine\ is 
one of the most N-enhanced stars yet found among CEMP stars ([N/Fe] $= +3.0$). 
In the following subsections we consider the nature of the light elements, O and 
other $\alpha$ elements, the odd-Z elements, the iron-peak elements, and the 
neutron-capture elements in our three program stars.

\subsection{ Li, C, N, and the $^{12}$C\,/\,$^{13}$C ratio}

We detect the Li~I 670.7 nm doublet in both \csthree\ and \cstwonine\ (see 
Fig. 1); A(Li) $\sim$ 1.7 is derived for both of these stars. The Li line 
is not detected in \cstwo; we derive an upper limit of A(Li) $< 0.6$ 
(3$\sigma$). It would clearly be of interest to obtain higher-S/N spectra 
of this star in order to strengthen this limit.

Both \csthree\ and \cstwo\ exhibit very large C enhancements; [C/Fe] 
$= +2.75$ and [C/Fe] $= +3.15$, respectively, as estimated from the average 
of results derived from the CH and C$_2$ features. In \cstwonine\ the C$_2$ 
lines are not detected; only the weak CH lines could be used to 
estimate [C/Fe] in this star. 

Nitrogen abundances for our stars are obtained from the CN 388.3 nm band (see Fig. 2); 
for \cstwo\ and \cstwonine\ we also use the NH band at 336 nm. The NH and CN 
lines yield slightly different abundances; we adopt the results from the CN 
lines, which have better S/N.

The $^{13}$CH lines at 401.9 nm (see Fig. 3) and the $^{13}$C, $^{12}$C 474.0 nm 
lines are used to derive $^{12}$C/$^{13}$C ratios. Only a weak
lower limit on the $^{12}$C/$^{13}$C ratio resulted for \csthree, while a
clear determination of $^{12}$C/$^{13}$C $= 9.0$ is found for \cstwo. The high
\teff and moderate C enhancement in \cstwonine\ make the $^{13}$C features 
very weak, and prevent us from placing any useful limits on $^{12}$C/$^{13}$C 
in this star.

\subsection{O and other $\alpha$ elements}

We estimate O abundances by LTE analysis of the O~I 
777.4 nm triplet, recognising that it is affected by both NLTE 
\citep{kiselman91,takeda,gratton,kiselman01} and 3D effects (Asplund 2005).
In \csthree\ the triplet lines are quite strong; we derive [O/Fe] =
+2.35. We do not detect the O~I lines in \cstwonine; hence we only
obtain an upper limit of [O/Fe] $<$ +1.4 (3$\sigma$). The [O/Fe] values in 
Table 3 include the NLTE corrections by \citet{gratton}. 

We find [Mg/Fe] $\sim +0.65$ in \csthree\ and [Mg/Fe]$ = +0.32$ and $+0.4$
in \cstwo\ and \cstwonine, respectively, similar
to most other VMP stars. Si is overabundant by +0.05 to +0.15
dex in \csthree\, and \cstwo, and Ca is overabundant in all three
stars by +0.2 to +0.4 dex. 

\subsection{The odd-Z elements}

Sodium abundances for our stars are determined using the Na~I D lines, which can
be affected by NLTE. We have applied the NLTE corrections by \citet{Bau98},
as listed in their Table 2.  The Na~I D lines in \cstwo\ are well separated 
from the significant interstellar component, 
due to the high radial velocity of this star. In \cstwonine\ the Na~I D1 line
merges with the interstellar Na I D2 line, so we derive the Na abundance from
the D2 line alone. Aluminium is underabundant in all three stars after an
NLTE correction of +0.65 dex \citep{Bau97}.

\subsection{The iron-peak elements}

The abundances of Ti, Sc, V, Cr, Mn, Co, and Ni are similar in all three 
stars, and close to the values reported for halo stars in the same metallicity 
range (e.g., Cayrel et al. 2004).

\subsection{The neutron-capture elements}

We derive strontium abundance estimates for our stars from the Sr~II 407.7 nm
and 421.5 nm lines. All three of our program stars exhibit Sr underabundances,
from $-0.2$ to $-0.4$ dex. A weak Y~II line at 377.4 nm is also detected in
\csthree, from which we derive [Y/Fe] $= -0.35$. Barium is moderately 
overabundant 
in both \csthree\ and \cstwonine, while it appears to be quite low in \cstwo.
Barium abundances were estimated from the Ba~II 455.4 nm and 493.4 nm lines. For 
\cstwo\ , we could not detect any Ba~II lines at all, and derive an upper limit 
of [Ba/Fe] $< -0.53$ (3$\sigma$), adopting the hyperfine splitting by 
\citet{mcw98} for Ba~II and assuming a solar isotopic composition.

\begin{table*}

\caption{Elemental abundances and $^{12}$C/$^{13}$C ratios for CEMP stars}
\begin{tabular}{lllllllllllll}

\hline
\hline
\scriptsize

Star         & \teff & \logg & [Fe/H]  & [C/Fe]& [O/Fe] & [N/Fe] & [Mg/Fe] & 
[Sr/Fe] &  [Ba/Fe] & $^{12}$C/$^{13}$C & A(Li) &  Ref \\
\hline

             &      &      &           &       &        &        &          &         
&         &       &         &      \\ 
{\bf CEMP-no stars$^{a}$} \\
             &      &      &           &       &        &        &          &         
&         &       &         &      \\ 
CS~22877-001 & 5100 &  2.2 &   $-$2.71 & +1.0  & \nodata& +0.0   & +0.29    & $-
$0.12 & $-$0.49 & $>$10 &  \nodata&    4 \\
CS~22949-037 & 4900 &  1.7 &   $-$3.79 & +1.17 & +1.98  & +2.57  & +1.56    &  
+0.33  & $-$0.58 &    4  &  \nodata&   10,17 \\
CS~22957-027 & 5100 &  1.4 &   $-$3.11 & +2.37 & \nodata& +1.62  & +0.69    & $-
$0.56 & $-$1.23 &    8  &  \nodata&   3,19 \\
CS~22958-042 & 6250 &  3.5 &   $-$2.85 & +3.15 & +1.12  & +2.15  & +0.32    & $-
$0.2  & $<$$-$0.53 &  9  &  $<$$0.6$&   TP \\
CS~29498-043 & 4600 &  1.2 &   $-$3.53 & +2.09 & +2.43  & +2.27  & +1.75    & $-
$0.47 & $-$0.46 &    6  &  \nodata&   3,7 \\
CS~29502-092 & 5000 &  2.1 &   $-$2.77 & +1.0  & \nodata& +1.0   & +0.37    & $-
$0.40 & $-$0.82 &  ~20  &  \nodata&    4 \\
G~77-61      & 4000 &  5.0 &   $-$4.0  & +3.2  & +2.2   & +2.2   & +0.49    & 
\nodata & \nodata & 5     &   $<$1.0&    18 \\
HE~0007-1832 & 6515 &  3.8 &   $-$2.65 & +2.55 & \nodata& +1.85  & \nodata  & 
\nodata & +0.16   & 10  &  \nodata&    9 \\
HE~0107-5240 & 5180 &  2.2 &   $-$5.3  & +4.1  & \nodata& +2.3   & \nodata  & 
\nodata & \nodata &\nodata&  \nodata&    8 \\
HE~1327-2326 & 6180 &  3.7 &   $-$5.6  & +4.1  & 2.8    & +4.5   & \nodata  &   
+1.1  & $<+1.5$ &\nodata&  $<1.5$&    11,11a,21 \\
             &      &      &           &       &        &        &          &         
&         &       &         &      \\ 
{\bf CEMP-no/s stars$^{b}$}  \\
             &      &      &           &       &        &        &          &         
&         &       &         &      \\ 
CS~29528-041 & 6166 &  4.0 &   $-$3.30 & +1.59 & $<$1.40& +3.07  & +0.4     & $-
$0.2  & +0.97   &\nodata&    1.71 &    TP \\
CS~31080-095 & 6100 &  4.5 &   $-$2.85  & +2.69 & 2.23  & +0.70  & +0.65    & $-
$0.41 & +0.77   & $>$40&    1.73 &    TP \\
             &      &      &           &       &        &        &          &         
&         &       &         &      \\ 

{\bf CEMP-s stars$^{c}$}  \\
             &      &      &           &       &        &        &          &         
&         &       &         &      \\ 
CS~22880-074 & 5850 &  3.8 &   $-$1.93 & +1.3  & \nodata& $-$0.1 &  \nodata &  
+0.39  & +1.31   & $>$40 &  \nodata&   3,19 \\
CS~22942-019 & 5000 &  2.4 &   $-$2.64 & +2.0  & \nodata& +0.3   &  \nodata &  
+1.7   & +1.92   &  8    &  \nodata&   3,19 \\
CS~30301-015 & 4750 &  0.8 &   $-$2.64 & +1.6  & \nodata& +0.6   &  \nodata &  
+0.3   & +1.45   &  6    &  \nodata&   3 \\
             &      &      &           &       &        &        &          &         
&         &       &         &      \\ 
{\bf CEMP-r/s stars$^{d}$}  \\
             &      &      &           &       &        &        &          &         
&         &       &         &      \\ 
CS~22898-027 & 6250 &  3.7 &   $-$2.26 & +2.2  & \nodata& +0.9   &  \nodata &  
+0.92  & +2.23   & $>$20 &    2.10 &   3,19 \\
CS~22948-027 & 4600 &  1.0 &   $-$2.57 & +2.43 & \nodata& +1.75  & +0.31    &  
+0.90  & +2.26   &  10   &    0.30 &   3,19 \\
CS~29497-030 & 6650 &  3.5 &   $-$2.70 & +2.38 & +1.67  & +1.88  & +0.64    &  
+0.84  & +2.17   &  $>$10&   $<$1.10 &   20 \\
             & 7000 &  4.1 &   $-$2.57 & +2.30 & +1.48  & +2.12  & +0.44    &  
+1.34  & +2.32   & \nodata &   \nodata &   13 \\
CS~29497-034 & 4800 &  1.8 &   $-$2.90 & +2.63 & \nodata& +2.38  & +0.72    &  
+1.00  & +2.03   &  12   &    0.10 &   12 \\
CS~29526-110 & 6500 &  3.2 &   $-$2.38 & +2.2  & \nodata& +1.4   &  \nodata &  
+0.88  & +2.11   &\nodata&  \nodata &   5    \\
CS~31062-012 & 5600 &  3.0 &   $-$2.55 & +2.1  & \nodata& +1.2   &  \nodata &  
+0.30  & +1.98   &  8    &  \nodata&   3,6 \\
CS~31062-050 & 5600 &  3.0 &   $-$2.33 & +2.0  & \nodata& +1.2   & +0.84    &  
+0.91  & +2.80   &  8    &  \nodata&   3,6 \\
HE~0024-2523 & 6625 &  4.3 &   $-$2.67 & +2.6  & +0.40  & +2.1   & +0.73    &  
+0.34  & +1.46   &  6    &    1.50 &   15 \\
LP~625-44    & 5500 &  2.8 &   $-$2.71 & +2.1  & \nodata& +1.0   &  \nodata &  
+1.15  & +2.74   &  20   &    0.40 &   1,2 \\
LP~706-7     & 6250 &  4.5 &   $-$2.74 & +2.15 & \nodata& +1.80  &  \nodata &  
+0.15  & +2.01   & $>$15 &    2.09 &   3,16 \\
HE~0143-0441 & 6370 &  4.4 &   $-$2.16 & +1.66 & \nodata& $-$0.04&  \nodata 
&\nodata  & +2.31   & 10    &  \nodata &   9 \\
HE~0338-3945 & 6160 &  4.1 &   $-$2.42 & +2.13 & +1.40  & +1.55  &  +0.30   &  
+0.74  & +2.41   &10     &  \nodata &   14 \\
HE~2148-1247 & 6380 &  4.3 &   $-$2.3  & +1.91 & \nodata& +1.65  &  \nodata & 
\nodata & +2.36   & 10    &  \nodata&   9 \\
             &      &      &           &       &        &        &          &         
&         &       &         &      \\ 
\hline

\end{tabular}
\noindent \parbox[t]{18cm}{
$^{a}$[C/Fe]$ > +1.0$ and [Ba/Fe] $<$ 0.0 (Beers \& Christlieb 2005)\\
$^{b}$[C/Fe]$ > +1.0$ and $+0.5 < {\rm [Ba/Fe]} < +1.0$ \\
$^{c}$[C/Fe]$ > +1.0$, [Ba/Fe] $>$ +1.0, and [Ba/Eu] $> +0.5$ (Beers \& 
Christlieb 2005)\\
$^{d}$[C/Fe]$ > +1.0$ and $0.0 < {\rm [Ba/Fe]} < +0.5$ (Beers \& Christlieb 
2005)\\

Ref.:
(1) Aoki et al. 2000;
(2) Aoki et al. 2001;
(3) Aoki et al. 2002a;
(4) Aoki et al. 2002b;
(5) Aoki et al. 2002c;
(6) Aoki et al. 2003;
(7) Aoki et al. 2004;
(8) Christlieb et al. 2002;
(9) Cohen et al. 2003;
(10) Depagne et al. 2002;
(11) Frebel et al. 2005;
(11a)Frebel et al. 2006;
(12) Hill et al. 2000;
(13) Ivans et al. 2005;
(14) Jonsell et al. 2006;
(15) Lucatello et al. 2003;
(16) Norris et al. 1997;
(17) Norris et al. 2002;
(18) Plez et al. 2005;
(19) Preston \& Sneden 2001;
(20) Sivarani et al. 2004;
(21) Aoki et al. 2006b;
(TP) This paper.}

\end{table*}

\section{Discussion}
 
Here we discuss the results for our three program stars in the context of
the CEMP class as a whole. Table 4 lists elemental abundances and 
$^{12}$C/$^{13}$C ratios for CEMP stars from the recent literature, grouped 
into the CEMP-no, CEMP-s, and CEMP-r/s classes. Our two moderately 
Ba-enhanced stars form a fourth group, which we call CEMP-no/s stars. 

We first discuss the C and N abundances for our CEMP stars. We next describe 
the nature of the oxygen abundance in these stars, followed by a 
discussion of Na, Mg, and other heavier elements. The variation of the 
$^{12}$C/$^{13}$C ratio with Li abundance is then considered. 

\begin{figure*}
\rotatebox{0}{\resizebox{16cm}{!}{\includegraphics{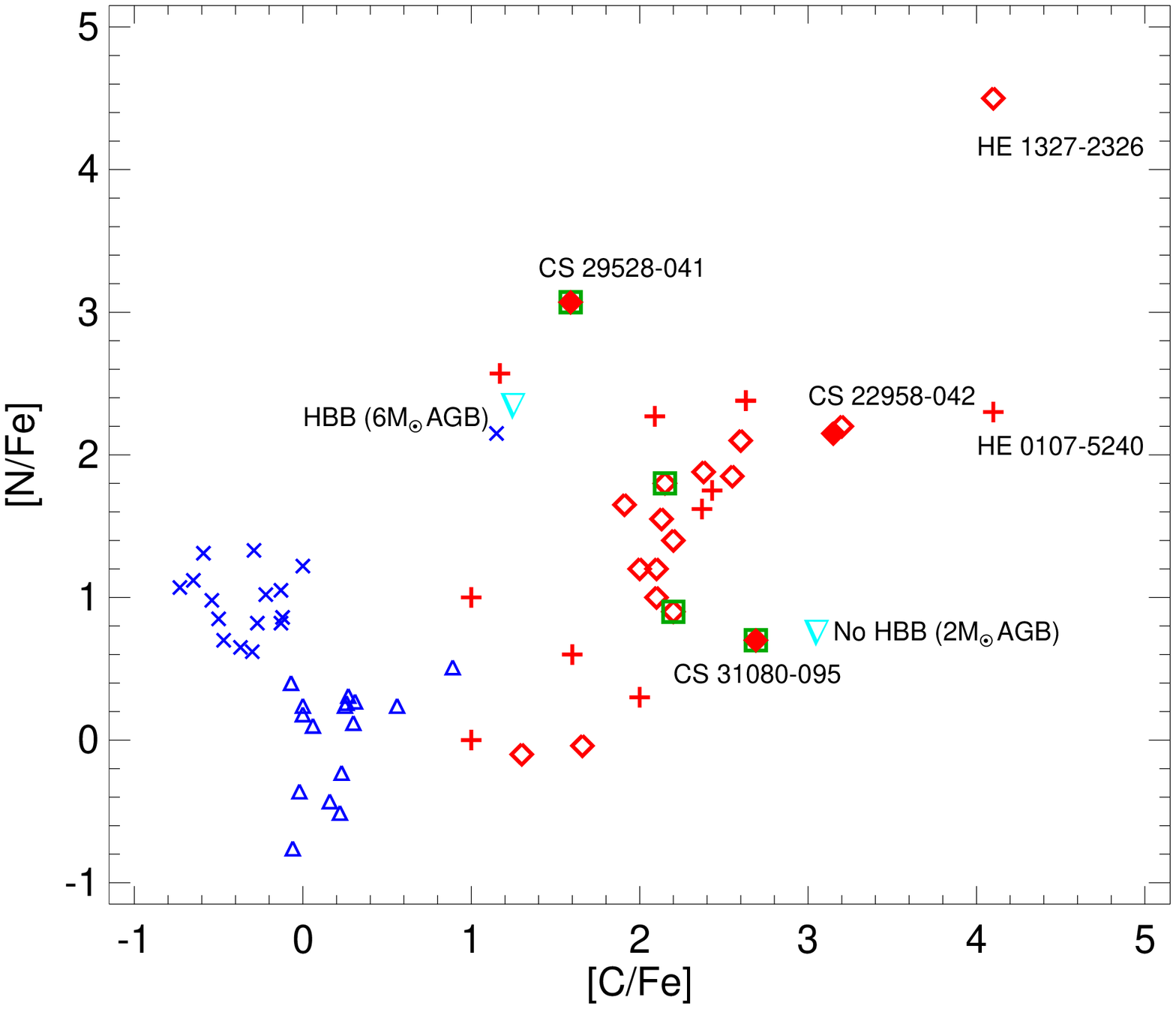}}}
\caption{ [N/Fe] vs. [C/Fe] for CEMP stars from the recent literature 
(Table 4). The $\times$ and $\triangle$ symbols represent the mixed and unmixed 
giants from Spite et al. (2005); the $+$ and $\diamond$ signs represent stars 
with \logg $<$ 3.5 and \logg $>$ 3.5, respectively. Stars with significant 
Li detections are surrounded by a $\sq$ (filled symbols: this paper). The 
AGB model predictions by Herwig (2004) for [Fe/H] = $-2.3$ and initial 
masses 6 \msun\ (HBB) and 2 \msun\ (non-HBB) are shown by $\triangledown$
symbols.  
HE~0107-5240 and HE~1327-2326 are the two known HMP stars.}
\end{figure*}

\subsection{C and N abundances in CEMP stars}

\csthree\ and \cstwo\ exhibit strong C and N enhancements, similar to most 
other CEMP stars.  \cstwo\ stands out as one of the most
carbon-rich of the known CEMP stars, while \cstwonine\ exhibits one of the
largest N abundances yet observed in CEMP stars.

Fig. 4 shows a broad overall correlation of the C and N overabundances
in most CEMP stars. Note that the ``carbon-normal'' stars from Spite et al. 
(2005), shown in the lower left part of the figure, 
display a clear anti-correlation between C and N due to mixing with CNO-cycled
material (see also Spite et al. 2006). 

C and N have distinct nucleosynthetic formation paths: C is made 
by He burning in the triple-$\alpha$ process, while N production requires 
H burning, i.e., two proton captures on \czw. Because the production site 
of primary C, the He-burning zone, is void of protons, primary C must
be mixed out of the He-burning region and into the H-rich outer
layers under conditions that are still hot enough for effective proton
capture. Thus, in a sense, it is not surprising to see high C and N together.

However, in stellar models it is very difficult to realize the specific
conditions for He and H burning and mixing that can produce the observed
abundance patterns quantitatively. Hot-Bottom Burning (HBB) in 
intermediate-mass stars has long been the only solid candidate
site for primary N production. Carbon is produced in the He-shell flashes and
dredged up into the envelope after the thermal pulse. During the 
interpulse phase between He-shell flashes the convective
envelope connects with the outer layers of the H-burning shell, and the C in the
envelope is transformed into N. 

As shown, for example, by the 4-6 $\msun$
models of Herwig (2004), the mass-averaged ejecta of such of VMP
intermediate-mass stars can reach [N/Fe]
$\sim +2$ to +3. Hot-bottom burning in intermediate-mass stars is characterized
by very efficient conversion of C and O into N. This finding is model
independent. The C overabundance in 4-6 $\msun$ VMP models is in the
range [C/Fe] $\sim +1$ to +2.

For stars that show no sign of the main component of the s-process or of being
binaries, one must consider other possible sources for the
observed C and N overabundances. Massive stars have been often dismissed
in the past because of the inability of standard models to produce primary N.
This inability is related to the above-mentioned necessity of mixing He-burning
products (C) out into the H-burning region. 

However, the recent models by Hirschi et al.
(2006) and Meynet, Ekstr\"om, \& Maeder (2006) show that massive, 
MMP ([Fe/H] $< -6.0$) stars that include the effect of rotationally-induced
mixing have the desired ability to produce large amounts of primary
N (which is required to match the observed behavior of [N/O] at low 
metallicity, even among ``carbon normal'' stars; see Chiappini et al. 2006). 
The C and N abundance signatures of rotating 
60 \msun\ models and the E-AGB envelopes of intermediate-mass stars (7 \msun) 
are quite similar (Meynet, Ekstr\"om, \& Maeder 2006).

C and N abundances from HBB models with initial mass 2 \msun\ and 
non-HBB AGB models with initial mass 6 \msun\ (Herwig 2004) are indicated 
in Fig. 4. Note
that because these (as well as the massive-star model predictions) are primary
yields, they shift along diagonals through zero in the [C/Fe]-[N/Fe] diagram as
a function of [Fe/H]. There is more than a one dex difference in [Fe/H] between
the HBB models and the non-HBB models. These two classes of models seem to 
delimit the region occupied by the CEMP stars. 

Most CEMP stars
in Figure 4 are located in an intermediate regime, showing larger N and smaller
C overabundances than predicted by the the 2 and 3 \msun\ AGB models, but less 
N and more C than predicted by the HBB AGB models. One of our program stars, 
\cstwo\ , falls into this category as well. Note, however, that \csthree\ 
exhibits [N/Fe] and [C/Fe] ratios in almost perfect agreement with the 
non-HBB models. 

Most other CEMP stars exhibit too strong N to be explained by standard 
low-mass AGB companions. It is interesting to note that the extra mixing, or 
cool-bottom burning, on the AGB that has been mentioned above in the context 
of Li abundances, could also increase the N abundance in 2 and 3 \msun\ 
VMP AGB stars, possibly to the extent required by observations. Unfortunately,
no quantitative models are currently available that consider this effect.

\cstwonine\ is one of the very few CEMP stars that exhibits the high N and 
low C expected from HBB AGB stars. Johnson et al. (2005, 2006) have searched for 
stars with low [C/N] ratios as predicted by current AGB models, using  
near-UV NH lines rather than the CN lines which may be weak in low-[C/N] 
stars; no obvious low-[C/N] candidates were found. 

This apparent paucity of N-rich EMP stars is a matter of concern. 
If the hypothesis that CEMP-s stars are mass-transfer binaries with AGB stars 
as donors is correct, then why do we not see the occasional Nitrogen-Enhanced 
Metal-Poor (NEMP) star that had a HBB AGB star as a companion? 
The C and N abundances suggest that \cstwonine\ is at least one candidate 
for these ``missing'' NEMP stars. 

\subsection{Oxygen}

The C and N overabundances of \csthree\ and \cstwo\ are typical of CEMP stars 
in general; however, they also exhibit significant O overabundances 
([O/Fe] = +2.35 and +1.35, respectively). Near solar metallicity, the production 
of O in AGB stars is negligible. It is therefore
at first sight surprising to see large overabundances of O in stars that are
considered to be substantially polluted by AGB stars. 

However, for [Fe/H] $\le -2$, AGB models produce primary O quite efficiently. 
In the model yields 
by Herwig (2004), the O overabundances range from [O/Fe] = +1.84 for the 
2 \msun\ case to [O/Fe] = +0.39 for the 6 \msun\ case. Again, because these 
are primary yields, they would to first order increase at lower [Fe/H]. 
Note that the O overabundance in the yields of Herwig (2004) should be 
a minimum, as they do not take convective overshoot into account; this 
would increase the O abundance in the AGB intershell region and 
increase the yields even further.

The O abundance in the He-shell flash convection zone is about 1-2\%
by mass, almost independent of metallicity. At solar metallicity this is 
about the initial abundance level. Thus, any primary O produced in the 
He-shell flash convection zone that is dredged up into the envelope will 
not significantly increase the surface O abundance. However, at extremely 
low metallicity, the envelope initially contains much less O  than the 
primary O contained in the He-flash intershell material that is 
dredged up into the envelope. Therefore, even a seemingly small O abundance 
of 1-2\% in the intershell region leads to a significant O excess compared 
to the initial O abundance. The existence of such an enhanced intershell 
O abundance is supported by observations of hot H-deficient central stars 
of planetary nebulae (Werner \& Herwig 2006).

For \cstwonine\, the low upper limit for O is consistent with HBB AGB
predictions. For \csthree\ and \cstwo\, the O abundance fits into the picture 
where the observed abundances are produced by non-HBB AGB stars.

\subsection{Sodium}

Sodium is extremely important for understanding the nuclear
production site of CEMP-star abundances. A small amount of secondary
Na can be made by C-burning in massive stars. However, with the large
overabundances of Na observed in \cstwo\ ([Na/Fe] = +2.82) and \cstwonine\ 
([Na/Fe] = +1.20) and many other CEMP stars, we need to look for a 
primary source of Na.

As explained above, primary production of $^{14}$N is difficult because
it requires the He-burning ashes to be exposed to  -burning again, which 
requires mixing. For Na this concept must be taken two steps
further. Once there is primary $^{14}$N from successive He and H burning,
the $^{14}$N must be brought back into the region of He burning in order
to capture two $\alpha$-particles. The result is primary $^{22}$Ne, 
which must then capture either a neutron or a proton
in order to make $^{23}$Na. Through the recurrent, interconnected He-shell
flash and dredge-up events, AGB stars provide a natural environment for
producing significant amounts of primary Na. In fact, AGB stars of all masses,
whether HBB or not, are predicted to produce primary Na.

Primary $^{14}$N exists in the ashes of the H-burning shell after a few
thermal pulses with dredge-up. This $^{14}$N is engulfed by the next He-shell
flash and transformed into $^{22}$Ne. In non-HBB AGB stars a neutron released
from the $^{22}$Ne($\alpha$,n)$^{25}$Mg reaction can be captured by $^{22}$Ne
again, leading to the production of $^{23}$Na; this $^{22}$Ne n-capture
is very efficient. In more massive AGB stars, the dominant effect is the
Ne-Na cycle burning of dredged-up $^{22}$Ne during HBB. In any case, the
predicted AGB production range is [Na/Fe] $\sim +1.0$ to $\sim +1.7$. Again,
since this is a primary production, this number would to first order 
increase at lower [Fe/H].

Hence, while rotating massive stars have been shown to be able to mix primary C
out of the core into the H-burning shell and thus produce primary N, Na
production requires that this N be exposed to He-burning once again. This may be
difficult to achieve in massive stars. 

Sodium is very low in \csthree, which poses a problem if one interprets
its CNO abundance as a result of AGB pollution. The low Na abundance
in \csthree, together with its high Mg and low Al abundance, does in fact fit 
the well-known odd-even pattern that is characteristic of massive-star yields
at low metallicity. The other two stars show this odd-even pattern starting at 
Mg; apparently, some additional primary source of Na filled in the 
low Na expected from the standard odd-even effect.

\begin{figure}
\rotatebox{0}{\resizebox{8.5cm}{!}{\includegraphics{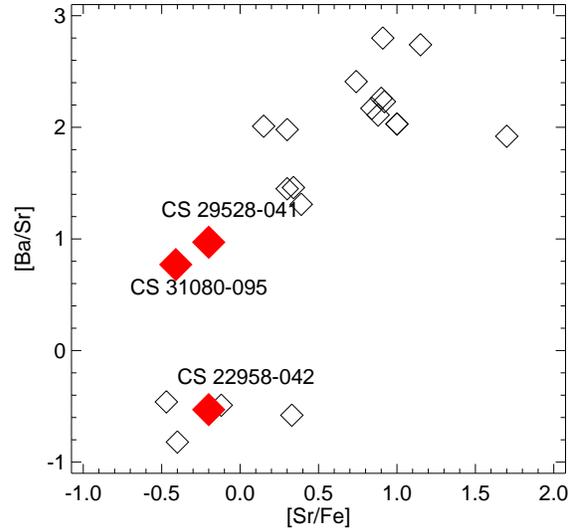}}}
\caption{[Sr/Fe] vs. [Ba/Sr] for the CEMP stars in Table 4 (filled 
symbols: this paper). [Ba/Sr] ratios  for
\csthree\ and \cstwonine\  falls in between CEMP-s stars and CEMP-no stars.
}
\end{figure}

\subsection{Neutron-capture elements}

\begin{figure*}
\rotatebox{-90}{\resizebox{12cm}{!}{\includegraphics{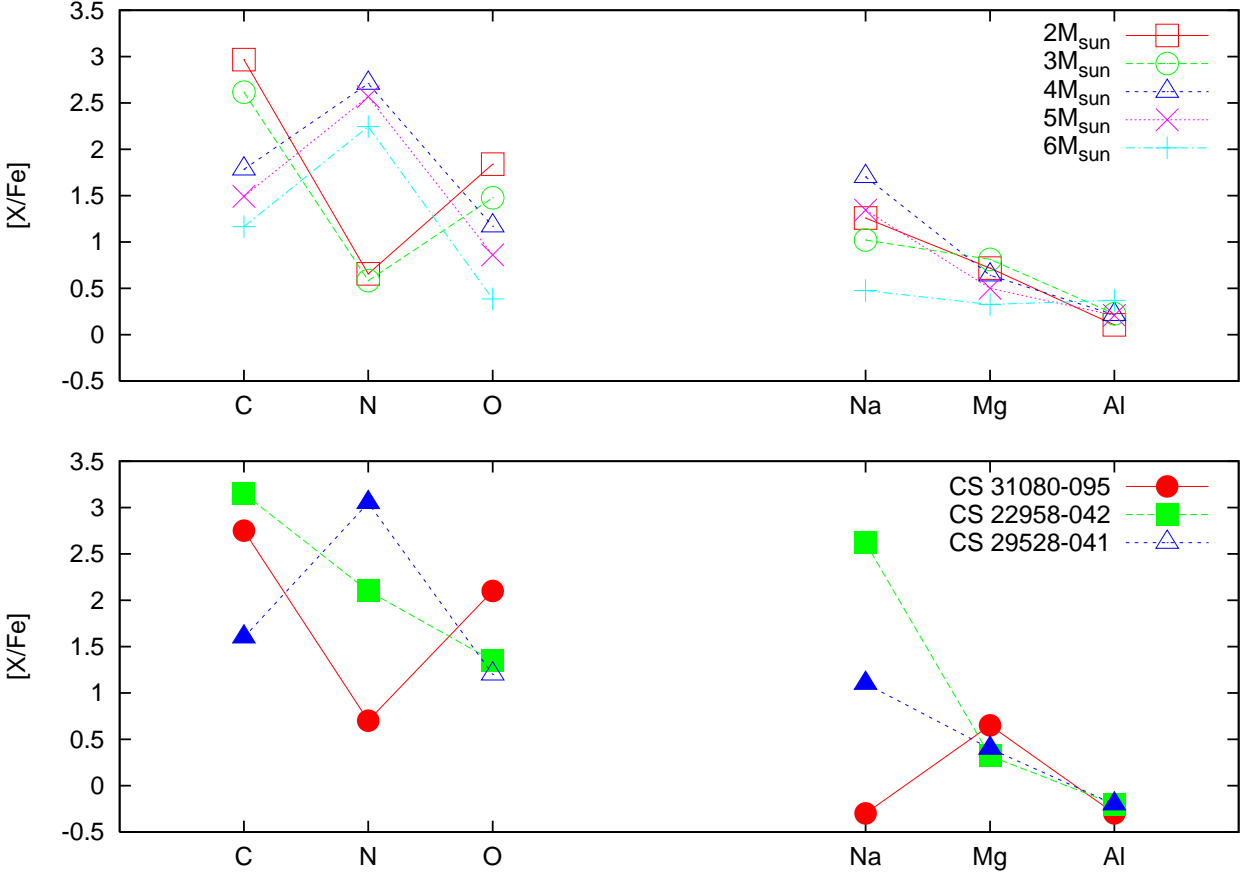}}}
\caption{{\it Top:} Element enhancements in VMP AGB model ejecta ([Fe/H] 
= $-2.3; Z = 0.0001$). {\it Bottom:} Observed abundances in our stars 
(open symbols: upper limits).  The abundances of \cstwonine\ agree well 
with HBB model predictions for an AGB mass of 4-6 \msun, but the other two 
stars are better matched by lower-mass AGB models. However, the observed 
high N abundances require enhanced mixing, and the observed Na may differ
more than expected from current models.}
\end{figure*}

The first s-process-peak elements Sr and Y are low in our stars 
(Y is detected only in \csthree), at the level seen in field giants of
similar metallicity \citep{francois06}. The second-peak 
s-process element Ba is moderately overabundant in \csthree\ and
\cstwonine; these two stars defy a clear classification into either
CEMP-s or CEMP-no, as can be seen in Fig. 5. All the CEMP-no stars 
listed in Table 4 are mildly C-rich and have much lower Ba than our two stars. 
At very low metallicities there might be an r-process component of Ba, but
Sr should then be higher than observed in our program stars. Accurate 
heavy-element abundances for a larger number of CEMP stars is
clearly needed to determine the origin of the heavy elements. 

In Table 4, \csthree\ and \cstwonine\ define a new class, CEMP-no/s, with 
properties intermediate between the CEMP-no and CEMP-s classes. In contrast,
CS~22949-037 \citep{depagne} displays a moderate Sr enhancement and low Ba
abundance.  

\csthree\ and \cstwo\ exhibit exactly opposing behaviour with
regard to Ba and Na. The interpretation of this result is not clear.  
The C/N ratio of both stars indicate that they carry the signature 
of  non-HBB AGB stars of low mass. Models predict qualitatively that 
for  non-HBB AGB stars Ba and Na should both be enhanced. In particular
the large Na and small Ba abundance of \cstwo\  means most of the Na  has not 
been produced by neutron captures on $^{22}$Ne in the He-shell  flash. 
However, as explained above, neutron captures are the source  Na in low-mass EMP 
stellar models. Compared to \csthree\   \cstwo\  has also a larger N/O abundance
than the low-mass stellar evolution  models (cf. Fig. 6). This indicates that 
the excess Na in \cstwo\ has  orinated in p-captures, possibly in a cool-bottom
burning type extra- mixing. It is conceivable that tidal interactions in the
binary  system have induced a variation of mixing, that could explain the  
difference between \csthree\ and \cstwo.

\subsection{Li in CEMP stars}

Few CEMP stars presently have spectra of sufficient quality from which
abundances or upper limits on A(Li) can be termined. For CEMP-s stars the source
of this Li is most likely a former AGB companion star, now a white dwarf. While
evolving through the giant phase, low- and intermediate-mass stars are expected
to deplete their initial Li (e.g. CS~22948-027, CS~29497-034, and LP~625-44).
Cool dwarfs should also exhibit low Li, due to convection-driven depletion in the
star itself (e.g., G~77-61).

Lithium should be preserved in warm dwarfs ($T_{eff}>$ 5700K). The existence of stars
with A(Li) slightly below the Spite Plateau (e.g., \csthree\ and \cstwonine)
indicates that some Li has been destroyed due to mixing processes in the 
progenitor(s) that produced the C (and N) enhancement. Some other CEMP stars, 
e.g., \cstwo\ or CS~29497-030, have values of \teff and \logg such that Li 
should not have been depleted; yet A(Li) is considerably below the Spite 
Plateau. While stars with Li slightly below the Spite Plateau can be 
explained by transfer of Li-depleted material from an AGB companion, a 
star like \cstwo\ must have been Li-poor already before this event.

Finally, the two CEMP-s stars LP~706-7 and CS~22898-027 
have Li abundances corresponding to the Spite Plateau. This implies that the
donor star has {\it produced} some Li, in an amount finely tuned to keep 
Li close to the Spite Plateau value on the companion star we now observe. 
A similar process has been suggested for the main-sequence turnoff stars 
in the globular cluster NGC 6397 \citep{bonifacio2002}.

Several mechanisms can in fact produce Li both during the Red Giant Branch 
(RGB) and AGB phases. On the RGB, a mechanism invoking 
enhanced, rotationally-induced mixing has been proposed and explored by
Denissenkov \& Herwig (2004). This is a particularly effective variant
of the cool-bottom processing proposed to explain some of the abundance
anomalies in corundum grains and globular clusters (Wasserburg, Boothroyd, \&
Sackmann 1995; Boothroyd \& Sackmann 1999). 

Enhanced mixing on the RGB can produce Li in a star that has 
depleted all its primordial Li, but
it cannot produce the large overabundances of primary C, N, and O that are
observed in the CEMP stars. However, if enhanced mixing is possible, and 
observed in RGB stars, it may be possible on the AGB as well, but  
it has not been yet been considered in any evolution model for VMP AGB stars.
The only quantitative treatment of such an evolution scenario was
by Nollett, Busso, \& Wasserburg (2003), who focused on isotopic ratios in solar
metallicity AGB stars, but not on Li. We may, nevertheless, speculate that dwarf
CEMP-s stars with detectable Li may be the result of enhanced mixing
during the AGB phase of their companions. 

We cannot exclude this origin for Li in \csthree. The fact that 
enhanced-mixing RGB models predict a low $^{12}$C/$^{13}$C ratio does not 
necessarily disagree with the high $^{12}$C/$^{13}$C measured in
\csthree, because the enhanced mixing may occur only in a transitional phase,
followed (or preceded) by efficient third dredge-up of primary $^{12}$C. Third
dredge-up will increase the $^{12}$C/$^{13}$C ratio, but in low-mass AGB stars
Li could still survive, at least partially. Clearly, a detailed theoretical
investigation of this scenario is highly desirable.

Another mechanism to produce Li in AGB stars is the well-known HBB
process, which operates in more massive AGB stars. The lower mass
limit for HBB decreases with declining metallicity and is about 3.5 $\msun$
for [Fe/H] $= -2.3$. Li production by HBB and efficient transformation 
of primary C into N are robust predictions of intermediate-mass
stellar evolution models; \cstwonine\ seems to fit this pattern.

Finally, a number of studies of the evolution of EMP, UMP, or 
zero-metallicity AGB stars have invoked the so-called H-ingestion flash 
to account for the observed abundance patterns in some of the
more extreme CEMP stars (Fujimoto, Ikeda, \& Iben 2000; Herwig 2003; Iwamoto et
al. 2004; Suda et al. 2004). The H-ingestion flash can also be a source of Li, 
as was first described by Herwig \& Langer (2001)
in the context of the born-again post-AGB stars (Werner \& Herwig
2006), and confirmed by Iwamoto et al. (2004) for VMP stars and by Suda et al.
(2004) for EMP and HMP AGB stars. 

Unfortunately, quantitative abundance 
predictions for this event remain very uncertain
because of our poor understanding of the coupled mixing and burning processes
that are involved. Probably, a H-ingestion flash that produces Li is
followed by several thermal pulses with efficient $^{12}$C dredge-up, which
would also make this scenario a candidate for explaining the observed Li
abundance in \csthree.
 
\subsection{$^{12}$C\,/\,$^{13}$C ratios}

\begin{figure}
\rotatebox{0}{\resizebox{8.5cm}{!}{\includegraphics{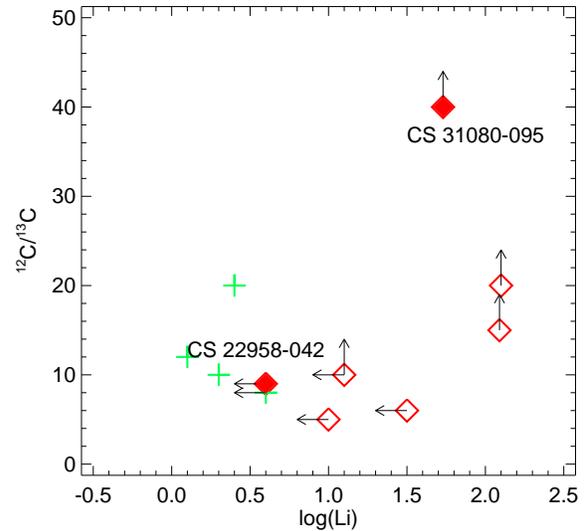}}}
\caption{$^{12}$C/$^{13}$C vs. A(Li) for the CEMP stars in Table 4 
(symbols as in Fig. 4; arrows indicate limits). The Li-rich CEMP stars 
tend to have high $^{12}$C/$^{13}$C ratios.
}
\end{figure}

Most CEMP-s stars exhibit carbon isotope ratios in the range
$^{12}$C/$^{13}$C $\sim$ 10-20, but the CEMP-no stars, including
\cstwo, have $^{12}$C/$^{13}$C $\sim$ 3-10 (CS~29502-092 is an exception).
Fig. 7 shows $^{12}$C/$^{13}$C vs A(Li) for CEMP stars with published 
detections or limits on these two quantities. $^{12}$C/$^{13}$C appears to 
be higher the in CEMP stars with detected lithium, but additional efforts 
to place tighter constraints on both $^{12}$C/$^{13}$C and Li in CEMP stars 
are clearly needed.

Standard VMP AGB star models predict that HBB produces $^{12}$C/$^{13}$C
$\sim 10$, while non-HBB models yield $^{12}$C/$^{13}$C of the order of 
$10^3 - 10^4$. The fact that no CEMP star in Table 4 (in particular no 
CEMP-s or CEMP-r/s star) has such a high $^{12}$C/$^{13}$C ratio is 
probably the clearest evidence that standard AGB models are neglecting some
mixing mechanism.  Even at solar metallicities the AGB models yield higher
$^{12}$C/$^{13}$C ratios than are observed, indicating the need for extra 
mixing (Charbonnel 1998). 
The extra-mixing process invoked above to account for the unexpectedly 
low C/N ratios in all CEMP-s stars is qualitatively suitable to produce 
the observed $^{12}$C/$^{13}$C ratios as well.

\section{Conclusions}

The origin of the observed abundance patterns of CEMP stars remains enigmatic.
Fig. 6 summarises how the abundances in our stars compare to the predicted 
abundances in AGB ejecta of very low metallicity, presumably transferred to 
the observed stars from former AGB binary companions. It should be
recognized that the stars have lower [Fe/H] than the models. Because
all the overabundances shown are dominated by the primary production in AGB
stars, this could be corrected to first order by just adding the Fe 
abundance difference (in logarithmic units). However, a stringent comparison 
of models and actual stellar observations must include the problems of 
mass transfer and dilution. For that
reason, one should not pay too much attention to absolute overabundances
of 0.5 -- 1 dex, but rather focus on the abundance patterns.

In our small sample, \cstwonine\ is probably the most straightforward case; 
all available evidence points toward pollution from a HBB AGB star. 
\cstwonine\ is one of the very few known NEMP stars, the existence of which is 
predicted by standard stellar evolution in conjunction with the binary
mass-transfer scenario for CEMP-s stars. The moderate Ba overabundance in this
star may be due to a polluting AGB star of large mass. The CNO abundances 
would also be compatible with pollution from a massive, rotating star. The 
s-process in these (or indeed any VMP AGB) stars is not yet understood. 
However, the observed Li, Na, and Ba abundances clearly favour the AGB
alternative.

Assuming that the measured rapid velocity variation for \cstwo\ is due to 
a binary orbital motion of  short period, the AGB connection strongly 
suggests that this star is a member of a post-common-envelope binary. 
This {\it must} be the case if the overabundances are attributed to
mass-transfer from an AGB star. 

We can only speculate about the effects of 
common-envelope phase on the nuclear signatures in a CEMP star
that formed from this mechanism. One scenario could involve several thermal
pulses with efficient dredge-up causing the observed overabundances of C, N,
O and Na. However, before the s-process production can get fully underway, AGB
evolution is terminated by the onset of common-envelope evolution. This could
explain the absence of Ba in this star. 

Another CEMP star with a short orbital period is HE~0024-2523 (Lucatello et al. 2003), which must be a
post-common envelope object as well. However, in that star Ba (and other heavy
elements) are overabundant without doubt. These two short-period binaries have
contrasting Na signatures: Na is underabundant in HE~0024-2523, but \cstwo\ is
one of the most Na-rich stars yet found.

Both HE~0024-2523 and \cstwo\ have very large C/O ratios, although O is
significantly overabundant: C/O$ = 48$ in \cstwo, and 100 in HE~0024-2523.
A very large C/O ratio may present a problem for the AGB mass-transfer scenario, 
because a non-HBB AGB star can only deliver material with a lower C/O ratio 
than found anywhere inside of it. The material dredged up to the surface, 
usually from the intershell, contains from one to a few percent primary 
O by mass, almost independent of metallicity, and also maybe $20 - 25\%$ 
carbon by mass\footnote{These C and O abundances are for models without 
extra convective mixing. If such mixing is included both C and O 
increase, but the C/O ratio does not change much.}.
Thus, C/O $<32$ is a reasonable limit for the intershell layers of AGB stars
(including VMP AGB stars). Even if the AGB star dumps this maximally
polluted material onto the CEMP star we now observe, the resulting
C/O ratio cannot become larger than in the intershell layer. The C/O ratio 
could be a problem for the two known short-period binaries, but perhaps 
especially for HE~0024-2523.

Clearly, common-envelope evolution introduces a whole new family of possible
evolutionary outcomes. Nevertheless, it is noteworthy that we may have found 
the second short-period CEMP binary, which seems to be significantly 
different from the previously known case, HE~0024-2523. One of the two might 
in fact not be a post-common envelope system, but have an inert low-mass MS
companion instead. However, VMP AGB stars remain the best present candidate 
nuclear production sites for explaining both the s-process enhancement
in HE~0024-2523 and the high Na in \cstwo.

Finally, \csthree\ could be an AGB mass-transfer object if VMP non-HBB AGB
stars experience some extra mixing. This could account for the observed Li, C,
and N abundances. The O abundance in this star is certainly high compared to the
predictions of this scenario, but cannot rule it out. The problem is the low
Na abundance, which excludes AGB progenitors above 2 \msun. 
Perhaps a less massive AGB star is involved, in which the $^{22}$Ne source is
not activated at all; the Ba could then still come from the $^{13}$C
neutron source. The larger $^{12}$C/$^{13}$C ratio ($> 40$) in this star is
consistent with this picture, and inconsistent with the obvious alternatives;
the $^{12}$C/$^{13}$C ratios in HBB AGB stars ($^{12}$C/$^{13}$C $\sim
10$) as well as the wind ejecta of rotating massive stars ( $^{12}$C/$^{13}$C
$\sim 4$) are much lower than the observed limit.

This discussion has clearly shown that our current understanding of the origin 
of the abundance patterns in CEMP stars is still in its infancy. The large
overabundances of some elements in CEMP stars probe one or more important
primary nuclear production sites, including those that may have set the
stage for the following galactic chemical evolution. Yet, we remain far from
having identified even the main players in this game. Tasks to be addressed
include a better understanding of mixing in stars of all masses, and of 
the impact of binary evolution on nucleosynthesis.  A significant increase 
in the numbers of CEMP stars with accurate, detailed elemental abundances is 
also an obvious next step.

\begin{acknowledgements}
We thank the referee for the comments and suggestions, which has
considerably improved the paper.
We thank the ESO staff for assistance during all the runs of our Large 
Programme.
R.C., P.F., V.H., B.P., F.S. \& M.S. thank the PNPS and the PNG for 
their support.  P.B. and P.M. acknowledge support from the MIUR/PRIN 
2004025729\_002 
and P.B. from EU contract MEXT-CT-2004-014265 (CIFIST). T.C.B. and T.S. 
acknowledge
partial funding for this work from grants AST 00-98508, AST 00-98549, and AST 
04-06784, as well as from grant PHY 02-16783: Physics Frontiers Center/Joint
Institute for Nuclear Astrophysics (JINA), all from the U.S. National Science 
Foundation.  B.N. and J.A. thank the Carlsberg Foundation and the Swedish and 
Danish 
Natural Science Research Councils for partial financial support of this work.
F.H. acknowledges support from the LDRD program at LANL through grant  
20060357ER. 

\end{acknowledgements}

\bibliographystyle{aa}

\end{document}